**ORIGINAL ARTICLE**

# Detection of Retinal Blood Vessels by using Gabor filter with Entropic threshold

## Mohamed. I. Waly[1], [2], Ahmed El-Hossiny[3]

1 Biomedical engineering department, Cairo higher institute for engineering, computer sciences and management, Cairo, Egypt

2 Medical equipment technology – collage of applied medical sciences, Majmaah University, Majmaah, KSA

3 Biomedical engineering department, high institute of engineering Shock academy Cairo, Egypt

*Corresponding Author*
*Mohamed .I. Waly*
*email: m.waly@mu.edu.sa*

**Abstract**

Diabetic retinopathy is the basic reason for visual deficiency. This paper introduces a programmed strategy to identify and dispense with the blood vessels. The location of the blood vessels is the fundamental stride in the discovery of diabetic retinopathy because the blood vessels are the typical elements of the retinal picture. The location of the blood vessels can help the ophthalmologists to recognize the sicknesses prior and quicker. The blood vessels recognized and wiped out by utilizing Gobar filter on two freely accessible retinal databases which are STARE and DRIVE. The exactness of segmentation calculation is assessed quantitatively by contrasting the physically sectioned pictures and the comparing yield pictures, the Gabor filter with Entropic threshold vessel pixel segmentation by Entropic thresholding is better vessels with less false positive portion rate.

**الملخص**

إعتلال الشبكية الذي يسببه مرض السكري هو السبب الأساسي لنقص البصر. هذا البحث يقدم استراتيجية مبرمجة لتحديد واستخلاص الأوعية الدموية داخل العين من الصور المأخوذة من قاع العين.

يعد تحديد موقع الأوعية الدموية هو خطوة أساسية في الكشف عن اعتلال الشبكية بواسطة مرض السكري لأن الأوعية الدموية هي العناصر النموذجية لصورة الشبكية. كما يمكن أن تساعد أطباء العيون بسرعة التعرف على الأمراض التي سبق ان اصابت العين.

تحديد الأوعية الدموية وا ستخلاصها تم باستخدام طريقة محرك الأقراص  والتحديق على قاعدتي الشبكية والتي يمكن الوصول إليها بحرية عن طريق مرشح جبير.

كانت نسبة خطا مرشح جبير بعد تعديل بعض العوامل أقل من قبل التعديل.

## I. Introduction

In the field of clinical ophthalmology, color retinal images obtained by a digital fundus camera, are broadly utilized for the detection and diagnosis of eye related diseases, hypertension, and numerous vascular disorders. Visual impairment is perhaps the most dreaded complication that might result from diabetes. This occurs because of problems present in the retina, resulting in an ailment known as retinopathy. Diabetic retinopathy is a disorder of the retinal vasculature. This condition sooner or later advances to some degree in almost all patients suffering from diabetes mellitus on the long term. The World Health Organization (WHO) states that the sum of adults with diabetes around the globe, would disturbingly escalate from a figure or 135 million in 1995 to a staggering 300 million in 2025 [1].

Computerized fundus picture investigation assumes a critical part in the PC supported analysis of ophthalmologic issue. A great deal of eye issue, and cardiovascular clutters, are known to relate to retinal vasculature changes. Numerous investigates has been done to investigate these connections. In any case, the clear majority of the examines depend on constrained information got utilizing manual or semi-computerized strategies because of the absence of robotized procedures in the estimation and examination of retinal vasculature. A fundus photo is a picture of the fundus taken by an ophthalmoscope, or a fundus camera. The primary pictures of the fundus were drawn by the Dutch ophthalmologist Van Trigt in 1853 [2]. An ordinary human fundus photo is appeared in Figure1. A typical human fundus photo is ruddy. Three noteworthy structures are displayed in an ordinary human eye fundus photo, the optic plate (or optic nerve head), the macula and veins. The optic plate gives off an impression of being a splendid locale in the picture, where all veins unite. Supply routes are brighter contrasting and veins. The macula is a dull locale where few veins show. Therapeutic signs, for example, exudates, hemorrhages, pigmentation, vein variations from the norm, and cotton fleece spots can be distinguished utilizing fundus photos. Fundus photography is still the most financially savvy picture methodology clinically. It is likewise regularly utilized as a part of screening projects, where the photographs can be examined later for conclusion and be utilized to screen the advance in numerous ailments. In a screening system, where a great many fundus photos are taken, it is unthinkable for the ophthalmologists to exam each photograph, particularly in the amazingly relentless work, for example, measuring vessel width for each vessel fragment. Semi-automatic methods have already been introduced to partially relieve the problem [3]. However, it is still insufficient, particularly in screening projects and some populace investigates where typically in any event several fundus pictures are included. The estimation and division of the retinal vasculature is of principle enthusiasm for the determination and treatment of various ophthalmologic conditions [4, 5].

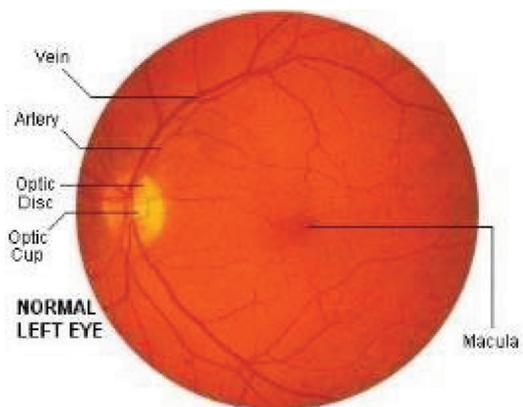

**Figure1. Normal Left Eye**

The exact segmentation of the retinal veins is every now and again an imperative precondition venture in the recognizable proof of retinal life systems and pathology. Moreover, the segmentation of the vessels is important for the recording of the patient pictures gained at different times [6-8]. Current vessel extraction techniques and algorithms can be categorized into four main methods as follows:

### a- Methods of matched filter

Matched filtering includes convolution of the images with array of filters for extraction of objects of interest. In [9], a 2-D kernel employed to segmentation of the vasculature. The filter shape is fabricated to counterpart that of a retinal blood vessel, which typically has a Gaussian derivative or a Gaussian profile. The kernel is usually rotated in 30 to 45 degrees increments to fit into blood vessels of diverse alignments. The maximum Matched Filter Response is chosen for each pixel and is normally threshold to generate a blood vessel image. As mentioned by many authors [10-12], a Matched Filter Response technique is effective when used in integration with supplementary processing techniques. Nevertheless, the convolution kernel may be quite large and needs to be used at different rotations causing a computational overhead, which may decrease the functioning of the whole segmentation method. In addition to that, the kernel responds perfectly to blood vessels that have the same standard deviation of the underlying Gaussian equation specified by the kernel. Thus, the kernel may not respond to blood vessels that have a dissimilar shape. The retinal low contrast of the smaller blood vessels and background variation also escalates the amount of false responses around bright objects, such as reflection artifacts and exudates. A lot of authors have suggested modifications and extensions that tackle the majority these difficulties [13-15]. In [13], the area and region based characteristics are used to detect the blood vessels in retinal fundus images. This technique observed matched filter response image using a probing technique. The probing technique classified pixels in a region of response image as vessels and non-vessels, by iteratively lessening the threshold. In each of the iteration, the probe examined the region-based attributes of the pixels in the tested region and detected the pixels classified as vessels.

### b- Methods of Vessel tracking

Vessel tracking technique detect a blood vessel between two points [16-19]. Different to the formerly described techniques for blood vessels detection, they work at the level of a single blood vessel rather than the full blood vessel. The vessel tracking technique characteristically steps along the vessel. Here, the center of the longitudinal cross-section of the blood vessel is determined with different characteristics of the blood vessel including average width and tortuosity measured during tracking the blood vessel. The main benefit of vessel tracking methods lies in the fact that they offer highly accurate blood vessel widths, and can gives information about each blood vessels that is usually impossible using other methods. However, these methods required the starting point of the blood vessel, and usually the end to be located by a user, and are thus, without additional methods, of limited use in fully automated analysis. In addition, vessel-tracking methods may be confused by vessel bifurcations and crossings and often lean towards termination at branch end points.

### c- Methods of Classifier based

The neural networks method has been widely examined for segmenting retinal structure such as the vasculature [20]. The procedure of a neural network is equivalent to that of a matched filter. Each method take sub windows of the image as input and return a chance measure as output. Two different researches, each one using the back-propagation approach, have detected [21] and segmented [22, 23] the retinal vasculature. Detection involves classifying sub windows as comprising vessels or no vessels. Segmentation include classification of each vessels and non-vessels pixels. In [22], the images are pre-processed with a principal component analysis to remove noise background. This takes place by reducing the dimensionality of the data set and then by applying a neural network to identify the pathology. They reported general specificity and affectability of 91% and 83.3%, individually. The result of the technique was contrasted and an accomplished ophthalmologist physically mapping out the area of the veins in an arbitrary example of seventy-three 20×20 pixel windows and requiring a correct match between pixels in both pictures. The neural networks researched by [21] used 20×20-pixel sub windows. Nine thousand of these sub windows were marked for neural learning validation. Generalization assessment over 1200 unseen sub windows resulted in a sensitivity of 91.7%. One of the benefits that made neural networks attractive in divide object into pieces of digital medical image is its capability to use nonlinear classification boundaries obtained during the training of the network and capability to learn. However, the main disadvantages of the neural networks are that they need to be trained every time whenever it introduced to a new feature to the network, and another limitation is the necessity for constructing the network with a gold standard or training data. The gold standard data is consisting of several binary images whose blood vessels must be precisely segmented by a professional ophthalmologist. However, as mentioned by [13], there is important disagreement in feel empathy of the blood vessels even amongst expert observers.

### d- Methods of Morphology

Morphological image processing technique utilizes features of the blood vessels morphology which called priori, such as it being piecewise linear and connected [24-26]. Calculations that concentrate straight shapes can be exceptionally significant for vessel division. A vessel division calculation from retinal angiography pictures in view of scientific morphology and straight preparing was offered [26]. An exceptional element of the calculation is that it utilizes a geometric model of all conceivable undesirable examples that could be gone head to head with vessels to separate vessels from them. The quality of the calculation originates from the gathering of numerical morphology and differential administrators in the division procedure. In [27], direct brilliant shapes and essential elements are removed utilizing scientific morphology administrators and vessels are extricated utilizing arch separation and Laplacian filter. Applying morphological closing to assist in recognizing of veins in the automated grading of venous beading by filling in any holes in the skeleton outline of the vein created throughout the processing method. The major disadvantage of wholly depending upon morphological methods is that they do not make use of the known blood vessel cross-sectional shape. In addition, this method works very well in the status of normal retinal images with uniform contrast, but this method works badly with noise due to diseases within the digital retinal images of eye [28].

This paper presents the blood vessels detection techniques based on Gabor filter with vessel pixel segmentation by Entropic thresholding on the fundus images because it is very fast and requires lower computing power. Therefore, the system can be used even on a very poor computer system.

## II. Material and Method

The general flow chart for the blood vessels detection shown in Figure 2.

**Stage 1:** The digital color retinal fundus images mandatory for create an automatic system are obtained from two publicly available retinal databases which are named DRIVE [29] and STARE [13].

**The STARE Database:** The images in the STARE (Structured Analysis of Retina) database comprise of twenty-one retinal fundus slides and their ground truth. The images are digitized slides captured by a Top Con TRV-50 fundus camera with 35°-degree field of view. Each slide was digitized to yield a 605 x 700-pixel image with 24bits per pixel. Figure.3 demonstrates a sample of an image present in the STARE database.

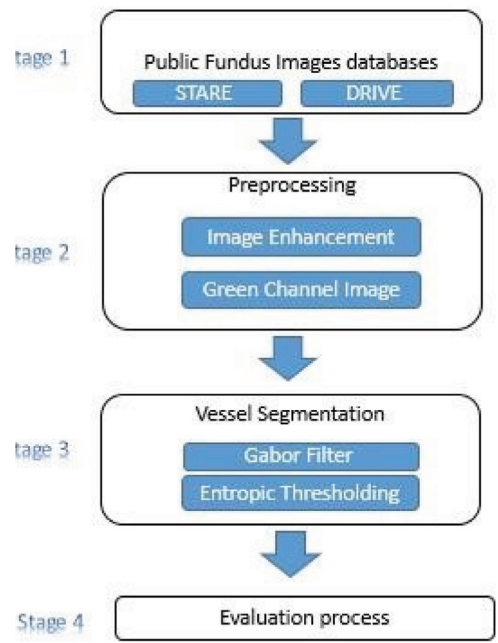

**Figure 2** **General flow chart for blood vessels segmentation**

**The DRIVE Database:** The DRIVE (Digital Retinal Images for Vessel Extraction) database, which is the second image database, comprises 40 color fundus photographs and their ground truth images. All the images in the DRIVE database are digitized using a Cannon CR5 non-mydriatic 3CCD camera with a 45° field of view. Each image is captured using 24-bits per pixel at the image size of 565×584. Figure.4 below shows an example of images in the database.

**Stage 2: Preprocessing**

Many factors can contribute towards causing a substantial quantity of images to be of an inferior quality. Such factors range from patient movement, poor focus, bad positioning, reflections, to inadequate illumination, as well as other similar factors, all which interfere with the analysis. Roughly 10% of the retinal images has objects that are weighty enough to obstruct human grading. Preprocessing of similar images can make certain achieve a satisfactory level in the automated detection of abnormalities. In the retinal images, there can be disparities that are instigated by numerous factors, including differences in cameras, illumination, acquisition angles, as well as retinal pigmentation. Initially during the preprocessing, the green channel of the retinal image is extracted, while subduing the other two color components, that is since

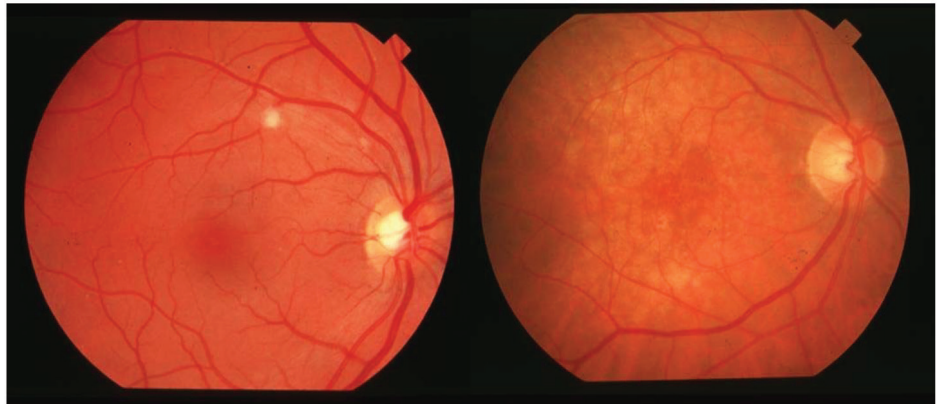
**Figure 3. Retinal image from STARE database.**

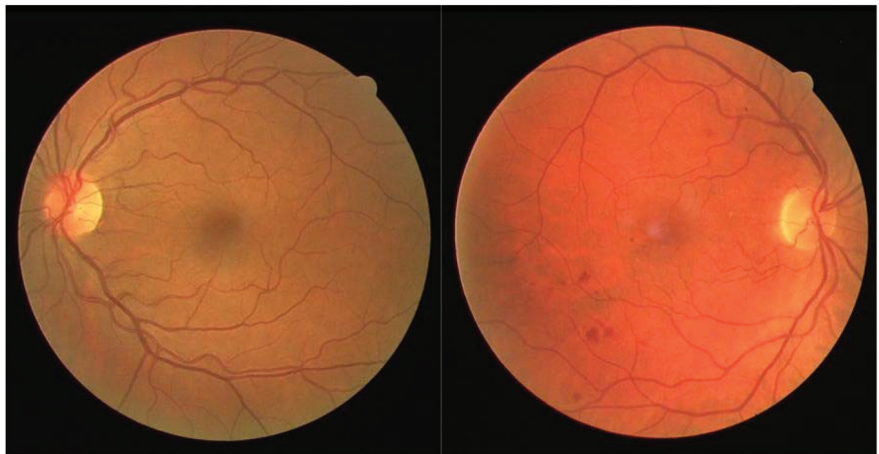
**Figure 4. Retinal image from DRIVE database**

most of the anatomical structure components of the retinal image appears most reliably in the green channel, as shown in figure 5. Few of the retinal images assimilated by utilizing standard clinical protocols repeatedly display low contrast. Moreover, retinal images characteristically have a higher contrast in the center of the image, with diminished contrast moving external from the center. For such images, a local contrast improvement technique is done as a second preprocessing step. Lastly, it is essential to produce a fundus mask for each image that is to enable the segmentation of lesions and anatomical structures in the following later stages. The preprocessing steps are detailed in the following subsections [30].

**Contrast Enhancement**

The contrast enhancement methods are intended to modify the visual appearance, to make an object distinct from other objects, and from the background. Typically, retinal images attained by the deployment of ordinary clinical protocols, show low contrast, and could comprise photographic items. Similarly, retinal image contrast lessened as the distance of a pixel from the center of the image escalates. In the current work, this preprocessing step is applied to the retinal images after the color normalization. Primarily, application of the histogram equalization on the intensity image produces.

It is realized that albeit the image quality is heightened, the central part of the image and the optic disc region are both over-enhanced, and this in turn causes the image to drop significant material. This is caused by histogram equalization characteristics that deals and affects the image overall. Since histogram equalization does not provide an effective structure, a Contrast-Limited

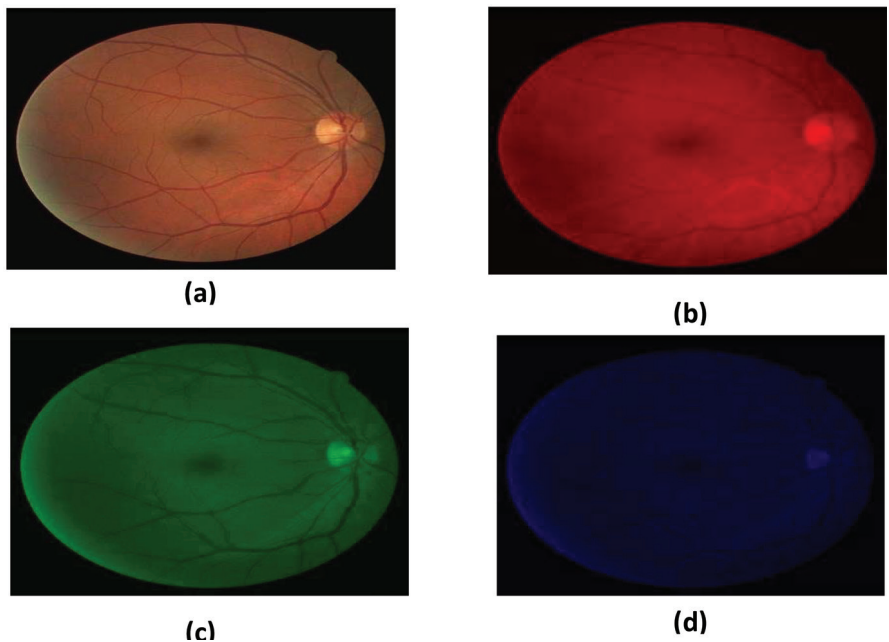

(a)

(b)

(c)

(d)

**Figure 5.** (a) Color retinal image; (b-d) Red, Green and image shown in figure 6.

Adaptive Histogram Equalization (CLAHE) technique was employed [22]. Whereas histogram equalization works on the whole image, CLAHE functions on small areas in the image, which called tiles. Each tile is contrast enhanced with histogram equalization. After performing the equalization, neighboring tiles are joined with the deployment of bilinear interpolation to eliminate artificially induced boundaries. While the contrast enhancement advances the contrast of exudate lesions, it also improves the contrast of some non-exudate background pixels, meaning that these pixels might be incorrectly branded as exudate lesions. For this reason, a median filtering operation is done to the intensity image prior to the contrast enhancement method, to reduce such an undesired effect [22].

**Fundus Mask Detection** The mask is a binary image, with the same resolution like that of a fundus image, whose positive pixels resemble the foreground area. It is central to distinguish between the fundus from its background, to facilitate further processing that is to be carried out only on the fundus, without interfering with those pixels belonging to the background. In a fundus mask, fundus pixels are marked as ones, while the background pixels are marked as zeros. The green channel image is threshold by a low threshold value, as the background pixels are normally considerably darker than the fundus pixels. A median filter of the size 5×5 is deployed to eliminate any noise from the created fundus mask, and the edge pixels are eradicated by morphological erosion with a structuring element of the size 5×5 also. Figure 7 shows an instance of the fundus mask.

**Stage 3: 2-D Gabor Filters for vessel enhancement.**

Gabor filters are straightforwardly connected to Gabor wavelets, as they can be designed for various dilations and rotations. Nevertheless, in general, expansion is not carried out for Gabor wavelets, due to it requiring the computation of bi-orthogonal wavelets, which might be extremely time-consuming. Consequently, more often a filter bank consisting of Gabor filters with various scales and rotations is generated. The filters involved with the signal, bringing about a so-called Gabor space. This process is thoroughly associated with processes in the primary visual cortex [31]. Jones and Palmer displayed that the real part of the complex Gabor function is a worthy fit to the receptive field weight functions found in simple cells in a cat's striate cortex [32].

The Gabor space is very helpful in image processing applications, such as optical character recognition, iris recognition, and fingerprint recognition. Associations amongst activations for a specific spatial location are very characteristic between objects in an image. Additionally, chief activations can be pulled out from the Gabor space facilitating the creation of a sparse object representation. The Gabor filters are sinusoidal modulated Gaussian functions that have ideal localization in mutually the frequency and space domains [33]. They are widely used within the computer vision community, with special consideration to its significance to the researches of the human visual system [34]. Gabor filters have been broadly deployed to image processing application problems, such as texture detection and segmentation [35], motion estimation, object detection, as well as strokes in character recognition [36], and roads in satellite image analysis [37,38]. The three foremost properties of the Gabor filters comprise first of their ability to be tweaked to match specific orientations, which permits the drawing out of the features of line segment at any possible orientation. Secondly, the bandwidth of the filter is modifiable. Thirdly and lastly, the output of the filter is strong to noise as it uses information of all the pixels within the kernel.

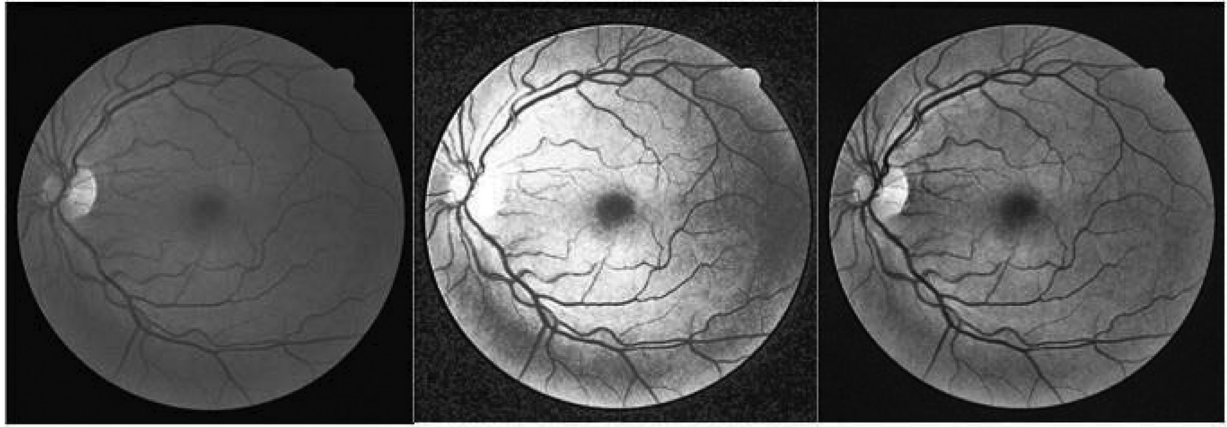

**Figure 6. (A) Green Channel, (B) after histogram equalization, (C) after CLAHE.**

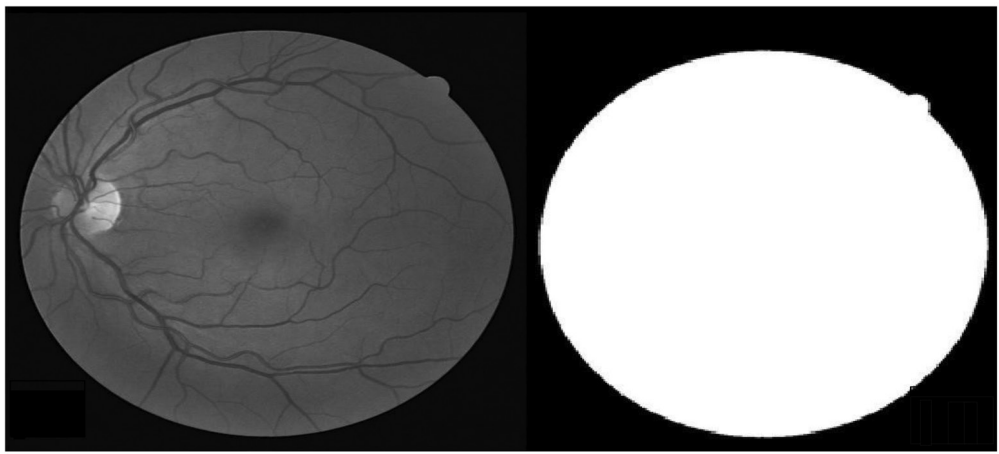

**Figure 7. Automatic fundus mask generation. (a) Input image; (b) Automatically generated fundus mask.**

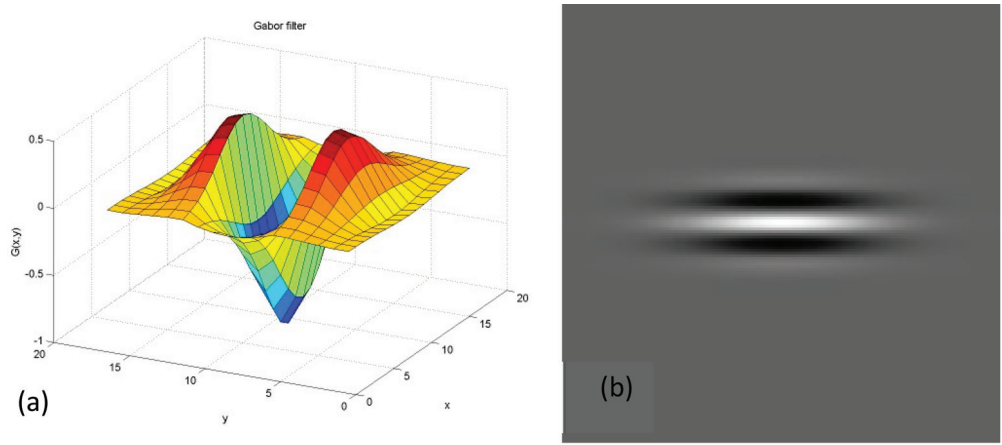

**Fig. 8: Gabor filter: (a) Surface representation, (b) Real part of filter.**

The actual part of a 2D Gabor filter is utilized in the context of retinal vessel segmentation that is defined in the spatial domain $g(x, y)$ as follows:

$$g(x, y) = \exp\left[-\pi\left(\frac{x_p^2}{\sigma_x^2} + \frac{y_p^2}{\sigma_y^2}\right)\right]\cos(2\pi f x_p) \quad (1)$$

Where,

$$x_p = x\cos\theta + y\sin\theta \quad (2)$$

$$y_p = -x\sin\theta + y\cos\theta \quad (3)$$

The parameters present in the Gabor function defined above are as follows. The angle $\theta$ is orientation of the filter, for example, a point of zero gives a channel that reacts well to the vertical elements in a picture. The parameter $f$ is the focal recurrence of pass band. Next, $\sigma_x$ is the standard deviation of Gaussian in x bearing along the channel that decides the data transfer capacity of the channel. At long last, $\sigma_y$ is the standard deviation of Gaussian over the channel that controls the introduction selectivity of the filter.

**Estimation of Gabor Parameters**

The size and shape of the lines or curvilinear structures to be identified are to be deliberated for deriving the Gabor parameters. In the case of vessel segmentation, the Gabor filter must be tweaked to a suitable frequency, so that vessels may be highlighted while background noise and other unwanted structures are filtered out. In [36], the procedure to attain the Gabor parameter values for the detection of straight line with width t has been shown. The same procedure is applied to estimate the Gabor parameters to distinguish vessels oriented in different directions.

The frequency $f$ determines the 2D spectral centroid positions of the Gabor filter. This parameter is derived with respect to the average width of the piecewise linear segments of retinal vessels and it is set to $f = \beta/t$ with $\beta$ set to a proper value in the range 0.5 to 1. $\lambda = \sqrt{2ln2/\pi}$ The standard deviation $\sigma_x$ determines the spread of the Gabor filter in the $\theta$ direction and it given by $\sigma_x = \lambda t/0.75\pi$. The standard deviation $\sigma_y$ that specifies the elongation of the filter is given by $\sigma_y = 0.85$ $\sigma_x$, where.

With the parameters set, the Gabor filter is thus said to be directional, as is its spatial domain, while support is in the form of an ellipse given by the elongated Gaussian of equation (1). Since, blood vessels have fluctuating diameters along diverse branches, suitable value of the thickness t must be established. When the value of t is large, most small vessels are dampened by neighboring noise. In manual segmentation, it is evident that majority of the vessel diameters are 120 $\mu m$ wide in a standard fundus image, with resolution of 20 micron/ pixels. Consequently, to provide lodging to all vessels in the image, the thickness parameter t is fixed at six, for the sake of enhancing and preserving of small vessels as well. The surface representation and real part of the resulting Gabor kernel is shown in Figure 8 with the angle $\theta$ set to zero. It is clearly apparent that it is appropriate for the orientation of directional features to provide a good response for pixels associated with the retinal blood vessels. It is also clear that the shape of the Gabor is locally like that of a blood vessel, and this shape is maintained across different orientations. Thus, more robust Gabor responses are formed when the filter is set up at the same position, orientation and scale as a vessel in the image.

Vessel enrichment is carried out on the input retinal image to highlight the vessel structures, while at the same time suppressing the background noise and other objects. As cited previously, the green channel in the RGB image offers the finest vessel to background contrast, when compared to the red and blue channels. Hence only the green channel image is utilized for more processing in the vessel segmentation method.

Due to the directional selectivity of the Gabor filter, enhancing the pixels of vessels oriented along various directions, with the parameters described earlier, is achievable. The response of applying a Gabor filter to a vessel segment is given by:

$$r\,(x, y) = g\,(x, y) * I_g\,(x, y) \quad (4)$$

Where, $g\,(x, y)$ is the Gabor filter defined by equation (1) and $I_g\,(x, y)$ is a green channel image with vessel segment oriented along diverse directions. The shape of the filter and the vessels are alike, and that when it is situated at the center of a vessel at scale and orientation, it offers a maximum response along the vessel direction, and minimum responses along its perpendicular direction. To identify the vessels oriented along different directions, the filter must be rotated among those directions, and only maximum response at that position is retained as follows:

$$r_{max}\,(x, y) = max_\theta\,[r\,(x, y)] \quad (5)$$

For each pixel position in an image, spatial filtering is achieved by convolving the image with the Gabor kernel along diverse orientations. It is also found in the literature that vessel segments lying within $\pm 7.5$ degree of the direction of the chosen kernel will respond well. So, the angle $\theta$ of the filter is rotated from 0 to 180 degrees in steps of $15^\circ$ to produce a single peak response on the center of a vessel segment.

**Segmentation of Vessels**

When the improved retinal vessels are compared to the background, the next step is to extract the vessels from the image. To extract the enhanced vessel segments in the Gabor filter response image in an appropriate manner, an effective thresholding scheme is obligatory. The entropy based thresholding using gray level co-occurrence matrix would be applied. This technique calculates the ideal threshold by considering the spatial spreading of the gray levels that are implanted in the co-occurrence matrix. From the co-occurrence matrix, several types of entropies such as global, local, joint and relative entropy, are computed to close the threshold value [39]. This technique is straightforward and uncomplicated to use since the co-occurrence matrix encompasses most of the information required for threshold value computation. For the appropriate segmentation of vessel pixels, a threshold is calculated using local entropy. The following section details the computation of the threshold.

**Gray Level Co-occurrence Matrix (GLCM)**

The GLCM contains information on distribute gray level frequency and edge information, as it is very useful in finding the threshold value [40, 41]. The gray level co-occurrence matrix is $L * L$ square matrix of the gray scale image I of spatial dimension $M * N$ with gray levels in the range $[0, 1 \ldots L - 1]$. It denoted by $T = [t_{i,j}]_{L*L}$ matrix. The elements of the matrix specify the number of transitions between all pairs of gray levels in a way. For each image pixel at spatial co-ordinate (m, n) with its gray level specified by f (m, n), it considers its nearest four neighboring pixels at the locations of (m+1, n), (m-1, n), (m, n + 1) and (m, n - 1). The co-occurrence matrix formed by comparing gray level changes of f (m, n) to its corresponding gray levels, f (m +1, n), f (m -1, n), f (m, n + 1) and f (m, n - 1). Depending upon the ways in which the gray level I follows gray level j, different definitions of co-occurrence matrix are possible. The co-occurrence matrix by considering horizontally right and vertically lower transitions is given by

$$t_{i,j} = \sum_{m=1}^{M} \sum_{n=1}^{N} \delta \qquad (6)$$

Where

$$\delta = 1 \; if \begin{cases} f(m,n) = i \; and \; f(m, n+1) = j \\ or \\ f(m,n) = i \; and \; f(m+1, n) = j \end{cases}$$
$$\delta = 0 \; otherwise$$

Normalizing the total number of transitions in the co-occurrence matrix, a desired transition probability $P_{i,j}$ from gray level I to gray level j is obtained as follows.

$$P_{i,j} = \frac{t_{i,j}}{\sum_{i=1}^{L} \sum_{j=1}^{L} t_{i,j}} \qquad (7)$$

**Entropic Threshold**

Based on the gray level variation within or between the object and the background, the gray level co-occurrence matrix is divided into quadrants. Let $T_h$ be the threshold within the range $0 \le T_h \le -1$ that partitions the gray level co-occurrence matrix into four quadrants, namely A, B, C and D.

In the figure 9, quadrant A represents gray level transition within the object while quadrant C represents gray level transition within the background. The gray level transition is between the object and the background or across the object's boundary placed in quadrant B and quadrant D. These four regions can be further grouped into two classes, referred to as local quadrant and joint quadrant. Local quadrant refers to quadrant A and C as the gray level transition that arises within the object or the background of the image. Then quadrant B and D are referred to as the joint quadrant because the gray level transition occurs between the object and the background of the image.

The local entropic threshold is calculated considering only quadrants A and C. The probabilities of the object class and background class are defined as:

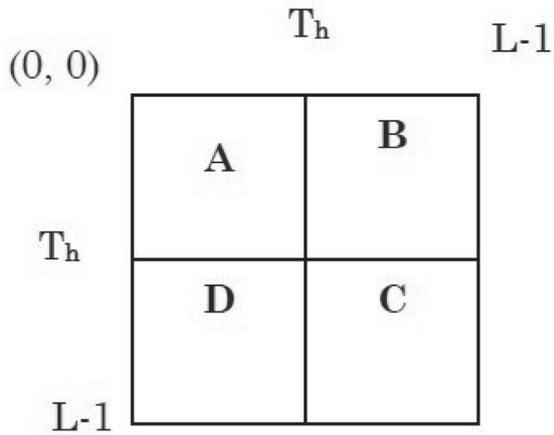

**Fig. 9: Four quadrants of co-occurrence matrix.**

$$P_A = \sum_{i=0}^{T_h} \sum_{j=0}^{T_h} P_{i,j} \qquad (8)$$

$$P_c = \sum_{i=T_h+1}^{L-1} \sum_{j=T_h+1}^{L-1} P_{i,j} \qquad (9)$$

Using equations (8) and (9) as normalization factors, the normalized probabilities of the object class and background class are functions of the threshold vector $(T_h, T_h)$, which is defined as:

$$P_{i,j}^A = \frac{P_{i,j}}{P_A} \qquad (10)$$

$$P_{i,j}^C = \frac{P_{i,j}}{P_C} \qquad (11)$$

From these, the local transition entropy A denoted by $H_A(T_h)$ and C denoted by $H_C(T_h)$ are calculated as follows:

The second-order entropy of the object given by:

$$H_A(T_h) = -\frac{1}{2} \sum_{i=0}^{T_h} \sum_{j=0}^{T_h} P_{i,j}^A \, log_2 P_{i,j}^A \qquad (12)$$

Similarly, the second-order entropy of the background given by:

$$H_C(T_h) = -\frac{1}{2} \sum_{i=T_h+1}^{L-1} \sum_{j=T_h+1}^{L-1} P_{i,j}^C \, log_2 P_{i,j}^C \qquad 13)$$

Both $H(T_h)$ and $H_C(T_h)$ are determined by the threshold $T_h$, thus they are functions of $T_h$. By summing up the local transition entropies, the total second-order local entropy of the object and the background is given by:

$$H(T_h) = H_A(T_h) + H_C(T_h) \qquad (14)$$

Finally, $T_E$, the gray level corresponding to the maximum of $H(T_h)$ over $T_h$ gives the ideal threshold for the value [42]:

$$T_E = arg\,[_{T=0...L-1} max\, H(T_h)] \qquad (15)$$

**Stage 4: Evaluation process**

The retinal images from the DRIVE database and STARE database are used for evaluating behave the vessel segmentation method. The manually segmented vessels provided in both the databases are used as a gold standard. The entire process of segmenting vessels was carried out on a Lenovo Idea Pad Z510 laptop with i7-4700MQ (2.40GHz, 6MB) CPU and 8GB memory using MATLAB 8.3. The processing of each image including convolution and thresholding took about 18 seconds.

A bank of twelve Gabor filters, oriented in the range of 0 to 180 degrees, are used to enhance the multi-oriented vessels. Increasing the number of filter banks did not result in significant improvement of the result, but increased the convolution operations. For each of the images, a corresponding manually segmented image is provided. Binary images with pixels are the ones determined to be part of a blood vessel by a human observer, who operated under instruct an ophthalmologist, and these images are colored white. Quantitative evaluation of the segmentation algorithm is carried out by comparing the output image with the corresponding manually segmented image. The comparison yields statistical measures that can be summarized using a contingency table, as shown in Table 1. True positives are pixels marked as vessels in both the segmentation given by a method and the manual segmentation used as ground truth. False positives are pixels marked as vessels by the method, but that are negatives in the ground truth. True negatives are pixels marked as background in both images. False negatives are pixels marked as background by the method, but are vessel pixels.

From these, sensitivity and specificity are assessed. Sensitivity gives the percentage of pixels correctly classified as vessels by the method, while specificity gives the percentage of non-vessels pixels classified as non-vessels by the method, they are calculated as follows:

$$sensitivity = \frac{T_p}{T_p + F_n} \tag{16}$$

$$specificity = \frac{T_n}{T_n + F_p} \tag{17}$$

Where $T_{op}$ is true positive, $T_N$ is true negative, $F_{op}$ is false positive and $F_{an}$ is false negative at each pixel. The method is compared with the matched filter based method of using the DRIVE database [15].

### III.    Results

An example of the quality of the obtained vessel boundary detection is shown in Figure 10 and 11. It shows a good performance on even the smallest vessels, which have low contrast. The proposed method was compared to the manual in Table 2 and Table 3.

**Table 1 Analysis using contingency table**

| | | Ground truth | |
|---|---|---|---|
| | | Positive | Negative |
| **Method result** | Positive | (True positive (T$_{op}$ | (False positive(F$_{op}$ |
| | Negative | (True negative(T$_{N}$ | (False negative(F$_{an}$ |

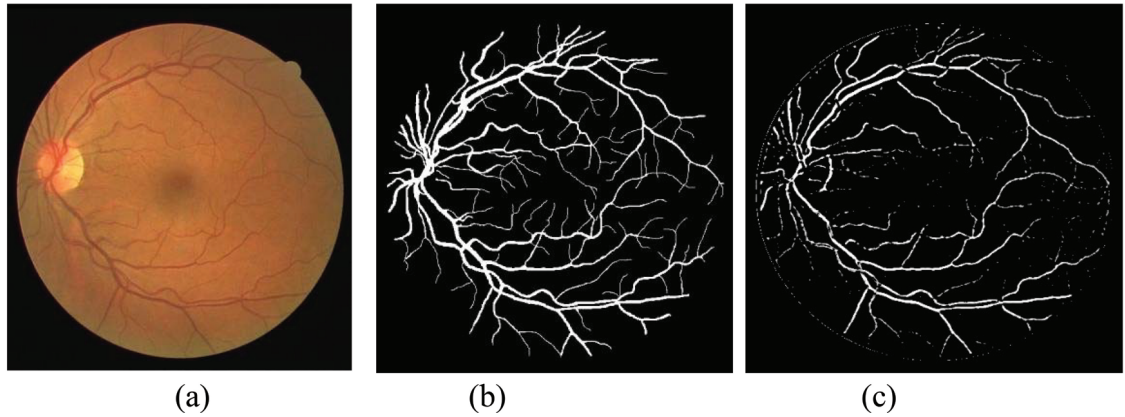

**Figure 10. (a) Original image from DIRVE (b) Manual segmentation (c) the proposed approach.**

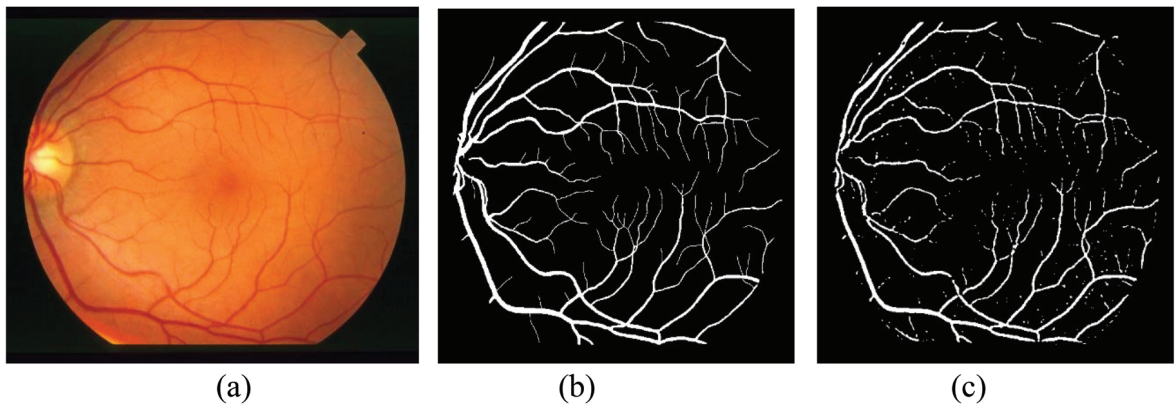

**Figure 11. (a) Original image from STAR (b) Manual segmentation (c) the proposed approach**

Table 2 shows that the Gabor filter is better in classifications of vessels with less false positive fraction rate.

The methods are also evaluated using the receiver operating characteristic (ROC) curve. ROC curves are formed by ordered pairs of true positive (sensitivity) and false positive (1-specificity) rates. The points on the ROC curve are attained by varying the threshold on the Gabor filter output image. For each configuration of threshold value, a pair formed by a true positive, and a false positive rate corresponding to the method's output, is marked on the graph figure 12. The closer an ROC curve is to the upper left corner, the better is behave the method, with the point (0,1) representing a perfect agreement with the ground truth. Consequently, an ROC curve is said to dominate another if it is completely above and to the left of it. In the experiment carried out, the ideal threshold

**Table 2 retinal blood vessels segmentation method on DRIVE database.**

| Method | Sensitivity (%) Mean ± SD | Specificity (%) Mean ± SD |
|---|---|---|
| Gabor filter | 4 ± 87 | 2 ± 96 |
| Matched filter | 2.62 ± 83.79 | 3.83 ± 89.59 |

used to get the last output varied in steps of 5 to obtain the number of points on the ROC curve for both the methods. In Figure, it can be obviously visualized that the Gabor filter method performs better than the matched filtered based method.

The results of the proposed method are also compared with those of [13], on twenty images from the STARE database and the result is shown in Table 2. Here it is evident that the proposed method also performs better with lower specificity, even in the presence of lesions in the abnormal images.

## IV.    Conclusion

This paper highlights signify obtain an inexpensive, noninvasive, highly safe, and easy to run tool and system that will be able to streamline the procedure of screening and diagnosing of diabetic retinopathy, as carried out by the physicians, via utilizing the digital image processing techniques to detect the main features of retinal fundus images, and some landmarks of diabetic retinopathy.

The main problems of the research were generalized the STARE database among its context, and sometimes the quality that affects directly the segmentation output. To solve this problem, we rejected 3 images from this database due to absence of several the anatomical structure, and then we applied the three image enhancement techniques to improve the quality of the images. The contrast limited adaptive histogram equalization method gave better results to advance the quality.

Ultimately, segmentations sequences executed based upon the literature review, experimental trials, and exiting packages. The segmentation stage was separated into four steps per the goal. To be able to verify the segmentation sequence, two verification

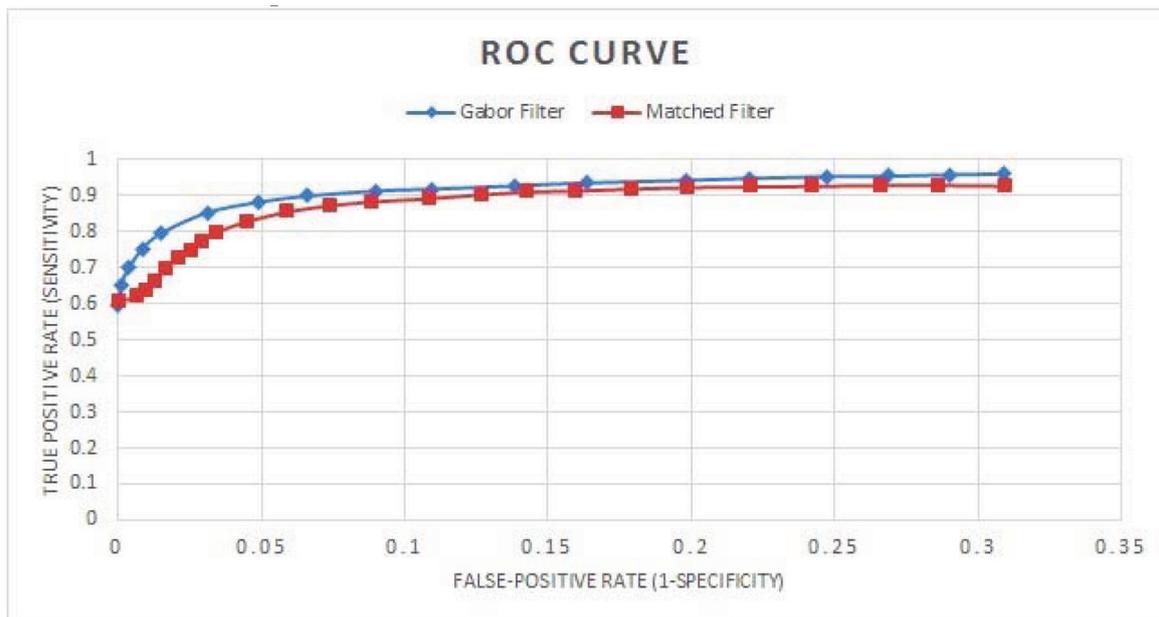

**Figure 12. ROC curve of Gabor filter.**

methods applied, a qualitative method as well as a quantitative method.

Gabor filters with entropic threshold have provided high-efficiency in the detection of retinal blood vessels. Large numbers of false-positive pixels observed around the optic nerve head: methods need to be developed to address this limitation. Further work is desired to design methods for ideal use of the information in the three-color components of the fundus images. The proposed methods could assist in the diagnosis of retinopathy.